\documentclass[aps,prb,twocolumn,groupedaddress]{revtex4-1}
\usepackage{hyperref}
\usepackage{amsmath}
\usepackage{graphicx}

\begin{document}

\title{Metal-insulator transition in three-band Hubbard model with strong spin-orbit interaction}
\author{Liang Du}
\affiliation{ Beijing National Laboratory for Condensed Matter Physics, 
              and Institute of Physics, 
              Chinese Academy of Sciences, 
              Beijing 100190, 
              China }

\author{Li Huang}
\affiliation{ Beijing National Laboratory for Condensed Matter Physics, 
              and Institute of Physics, 
              Chinese Academy of Sciences, 
              Beijing 100190, 
              China }

\affiliation{ Science and Technology on Surface Physics and Chemistry Laboratory, 
              P.O. Box 718-35, 
              Mianyang 621907, 
              Sichuan, 
              China }

\author{Xi Dai}
\affiliation{ Beijing National Laboratory for Condensed Matter Physics, 
              and Institute of Physics, 
              Chinese Academy of Sciences, 
              Beijing 100190, 
              China }

\date{\today}

\begin{abstract}
Recent investigations suggest that both spin-orbit coupling and electron correlation play very crucial roles
in the $5d$ transition metal oxides. 
By using the generalized Gutzwiller variational method and dynamical mean-field theory with the 
hybridization expansion continuous time quantum Monte Carlo as impurity solver, 
the three-band Hubbard model with full Hund's rule coupling and spin-orbit interaction terms, 
which contains the essential physics of 
partially filled $t_{2g}$ sub-shell of $5d$ materials, is studied systematically.
The calculated phase diagram of this model exhibits three distinct phase regions, including metal, band insulator and 
Mott insulator respectively. 
We find that the spin-orbit coupling term intends to greatly enhance the tendency of the Mott insulator phase.
Furthermore, the influence of the electron-electron interaction on the effective strength of spin-orbit coupling in the metallic 
phase is studied in detail. We conclude that the electron correlation effect on the effective spin-orbit coupling is far beyond 
the mean-field treatment even in the intermediate coupling region.
\end{abstract}

\maketitle

\section{introduction}
\label{sec:intro}

The Mott metal-insulator transition (MIT) induced by electron-electron correlation 
has attracted intensive research  activities in the past several 
decades\cite{Imada:RevModPhys.70.1039,Georges:RevModPhys.68.13,Kotliar:RevModPhys.78.865,Medici:PhysRevLett.107.256401}.
Although the main features of Mott transition have already been
captured by single-band Hubbard model, most of Mott transition in realistic materials have multi-orbital
nature and should be described by multi-band Hubbard model. 
Unlike the situation in single-band case, 
where the Mott transition is completely driven by the local
Coulomb interaction $U$, the Mott transition in multi-band case is affected by not only Coulomb interaction 
but also crystal field splitting and Hund's rule coupling among different 
orbitals\cite{Medici:PhysRevLett.102.126401,Werner:PhysRevLett.99.126405,kita:PhysRevB.84.195130}.
The interplay between Hund's rule coupling and crystal field 
splitting generates lots of interesting phenomena in the multi-band Hubbard model, 
for examples, orbital selective Mott transition, high-spin to low-spin transition and 
orbital ordering. Therefore, most of the intriguing physics in $3d$ or $4d$ transition 
metal compounds can be well described by the 
multi-band Hubbard model with both Hund's rule coupling and crystal field splitting.

In the present paper, we would like to concentrate our attention on the Mott physics in another group of interesting 
compounds, the $5d$ transition metal compounds,
where spin-orbit coupling (SOC), the new physical ingredient in Mott physics, plays 
an important role. Compared to $3d$ orbitals, the
$5d$ orbitals are much more extended and the correlation effects are not expected to be 
important here. 
While as firstly indicated in reference \cite{Kim:2008p4}, the correlation 
effects can be greatly enhanced by SOC, which is commonly strong in $5d$ materials. 
The first well studied $5d$ Mott insulator
with strong SOC is Sr$_2$IrO$_4$, where the SOC splits the $t_{2g}$ bands into (upper) $j_{\text{eff}}=1/2$ 
doublet and  (lower) $j_{\text{eff}}=3/2$ quartet bands and greatly suppresses 
their bandwidths\cite{Kim:2008p4,Pesin2010a,Kim:Science.329.1329,Watanabe:PhysRevLett.105.216410,Jackeli:PhysRevLett.102.017205}.
Since there are totally five electrons 
in its $5d$ orbitals, the $j_{\text{eff}}=1/2$ bands 
are half filled and the $j_{\text{eff}}=3/2$ bands are
fully occupied, which makes the system being effectively a $j_{\text{eff}}=1/2$ single-band 
Hubbard model with reduced bandwidth. 
Therefore the checkerboard anti-ferromagnetic ground state of Sr$_2$IrO$_4$ 
can be well described by the single-band Hubbard model with half filling. 
%

Here, we will focus on the $5d$ materials with four electrons in the $t_{2g}$ sub-shell. 
These materials include the newly discovered BaOsO$_3$,
CaOsO$_3$ and NaIrO$_3$ etc\cite{Bremholm2011601}. All these materials share one important common feature: 
in low temperature, these materials are insulators without magnetic long-range order. 
The origin of the insulator behavior can be due to two possible reasons, the strong enough Coulomb interaction and SOC.
We will have Mott insulator in the former and band insulator in the latter case respectively. Therefore it is interesting to study the features of 
metal-insulator transition in a generic $t_{2g}$ system occupied by four electrons with both Coulomb interaction and SOC.


In the present paper,
we study the $t_{2g}$ Hubbard model with SOC and four electrons filling by using
rotational invariant Gutzwiller approximation (RIGA) and dynamical mean-field theory combined with 
the hybridization expansion continuous time quantum Monte Carlo (DMFT + CTQMC) respectively.
The paramagnetic $U-\zeta$ phase diagram is derived carefully.
Further, the interplay between SOC $\zeta$ and Coulomb interaction $U$ is analyzed in detail. 
We will mainly focus on the following two key issues: (i) How does the SOC affect the boundary 
of Mott transitions in this three-band model? (ii) How does the Coulomb interaction modify 
the effective SOC strength? 

This paper is organized as follows. In Sec. \ref{sec:model}, the three-band 
model is specified, and the generalized multi-band Gutzwiller 
variational wave function is introduced. 
In Sec. \ref{subsec:rot}, the calculated results, including $U-\zeta$ phase diagram, 
quasi-particle weight and charge distribution, for 
the three-band model are presented. The effect of Coulomb 
interaction on SOC is analyzed in Sec. \ref{subsec:soc}. 
Finally   we make conclusions in section \ref{sec:conclusion}.

\section{model and method}
\label{sec:model}
%
%
The three-band Hubbard model with full Hund's rule coupling and SOC terms is defined by the Hamiltonian:
\begin{equation}
 H = - \sum_{ij,a\sigma}t_{ij}d_{i,a\sigma}^{\dagger}d_{j,a\sigma} + \sum_{i}H_{loc}^{i}
   = H_{kin} + H_{loc},
\label{eq1}
\end{equation}
where $\sigma$ denotes electronic spin, and $a$ represents the three $t_{2g}$ orbitals with 
$a=1,2,3$ corresponding to $d_{yz}, d_{zx}, d_{xy}$ orbitals respectively.
The first term describes the hopping process of electrons between
spin-orbital state ``$a\sigma$" on different lattice sites $i$ and $j$. 
Local Hamiltonian terms $H_{loc}^{i}=H_{u}^{i}+H_{soc}^{i}$ 
contain Coulomb interaction $H_u^i$ and SOC $H_{soc}^i$ 
(In the following, the site index is suppressed for sake of simplicity).
\begin{eqnarray}
 H_u & = &U    \sum_{a} n_{a\uparrow} n_{a\downarrow}
       + U'    \sum_{a<b,\sigma\sigma'} n_{a\sigma} n_{b\sigma'}
       - J_z   \sum_{a<b,\sigma} n_{a\sigma} n_{b\sigma} \nonumber\\
     && - J_{xy}\sum_{a<b} \left(
         d_{a\uparrow}^\dagger d_{a\downarrow} d_{b\downarrow}^\dagger d_{b\uparrow}
       + d_{a\uparrow}^\dagger d_{a\downarrow}^\dagger d_{b\uparrow} d_{b\downarrow} + h.c. \right),
\end{eqnarray}
\begin{equation}
 H_{soc} = \sum_{a\sigma}\sum_{b\sigma'} \zeta \langle a\sigma| l_x s_x + l_y s_y 
         + l_z s_z |b\sigma' \rangle d_{a\sigma}^\dagger d_{b\sigma'},
\end{equation}
where $U$ ($U'$) denotes the intra-orbital (inter-orbital) Coulomb interaction, 
$J_z$ term describes the longitudinal part of the Hund's coupling. While the other two 
$J_{xy}$ terms describe the spin-flip and pair-hopping process respectively. 
$\zeta$ is SOC strength, $l$($s$) is orbital (spin) angular momentum operator. 
%
%
Here we assume the studied system experiences 
approximately cubic symmetry ($O_h$ symmetry), in which two parameters $U'$ and $J_{xy}$ 
follow the constraints $U'=U-2J$ and $J_{xy}=J_z=J$. Here we choose $J/U=0.25$ for the systems studied in this paper unless otherwise noted. 
This lattice model is solved in the framework of 
RIGA\cite{Bunemann:PhysRevB.57.6896,Bunemann:PhysRevLett.101.236404,Nicola:PhysRevB.85.035133,Deng:PhysRevB.79.075114} 
and DMFT(CTQMC)\cite{Werner:PhysRevB.74.155107,RevModPhys.83.349} 
methods respectively, 
which are both exact  in the limit of infinite spacial dimensions\cite{PhysRevLett.59.121,PhysRevB.37.7382}.
In this work, a semi-elliptic bare density of states $\rho(\epsilon)=(2/\pi D) \sqrt{1-(\epsilon/D)^2}$ 
is adopted, which corresponds to Bethe lattice with infinite connectivity. In the present study, 
the energy unit is set to be half bandwidth $D=1$ and all orbitals are assumed to have equal bandwidth. 

%
%
Next, we will briefly introduce the recently developed RIGA method\cite{Nicola:PhysRevB.85.035133}. 
The generalized Gutzwiller trial wave function
$|\Psi_G\rangle$ can be constructed
by acting a many-particle projection operator $\mathcal{P}$
on the uncorrelated wave function $|\Psi_{0}\rangle$\cite{PhysRevLett.10.159,PhysRev.134.A923,PhysRev.137.A1726},
\begin{equation}
    |\Psi_{G}\rangle=\mathcal{P}|\Psi_{0}\rangle,
\end{equation}
with  
\begin{equation}
    \mathcal{P} = \prod_{i}\mathcal{P}_{i}=\prod_{i}\sum_{\Gamma\Gamma'}
          \lambda_{\Gamma\Gamma'}|\Gamma\rangle_{ii}\langle\Gamma'|.
\end{equation}
$|\Psi_{0}\rangle$ is normalized uncorrelated wave function in which
Wick's theorem holds. $|\Gamma\rangle_{i}$ are atomic eigenstates on
site $i$ and $\lambda_{\Gamma\Gamma'}$ are Gutzwiller variational parameters.
In our work, $|\Gamma\rangle_{i}$ are eigenstates of atomic Hamiltonian
$H_u$, each $|\Gamma\rangle_i$ is labeled by good quantum
number $\mathcal{N},\mathcal{J},\mathcal{J}_z$, where $\mathcal{N}$ is total number of electrons, $\mathcal{J}$ is total angular momentum, 
$\mathcal{J}_z$ is projection of total angular momentum along $z$ direction. The non-diagonal elements
of the previously defined variational parameter matrix $\lambda_{\Gamma\Gamma'}$ are
assumed to be finite only for state $|\Gamma\rangle$, $|\Gamma'\rangle$
belonging to the same atomic multiplet, i.e, with the same three quantum labels\cite{Bunemann:PhysRevLett.101.236404}.
In the following, we assume the local Fock terms are absent in $|\Psi_{0}\rangle$, 
\begin{equation}
   \langle\Psi_{0}|c_{i\alpha}^{\dagger}c_{i\beta}|\Psi_{0}\rangle
 = \delta_{\alpha\beta}\langle\Psi_{0}|c_{i\alpha}^{\dagger}c_{i\alpha}|\Psi_{0}\rangle
 = \delta_{\alpha\beta}n_{i\alpha}^{0}.
\end{equation}
For general case, a local unitary transformation matrix $\mathcal{A}$
is needed to transform the original $d_{i\alpha}$-basis into the so-called natural
$c_{im}$-basis\cite{Nicola:PhysRevB.85.035133}, i.e, $d_\alpha = \sum_m \mathcal{A}_{\alpha m} c_m$.
In the original single particle basis 
($d_{yz}\!\!\uparrow$,$d_{yz}\!\!\downarrow$, $d_{zx}\!\!\uparrow$,$d_{zx}\!\!\downarrow$, 
$d_{xy}\!\!\uparrow$,$d_{xy}\!\!\downarrow$),
SOC term is expressed as :
\begin{eqnarray}
    H_{soc} = - \frac{\zeta}{2}
    \left(\begin{array}{cccccc}
    0  &  0  & -i  &  0  &  0  &  1  \\
    0  &  0  &  0  &  i  & -1  &  0  \\
    i  &  0  &  0  &  0  &  0  & -i  \\
    0  & -i  &  0  &  0  & -i  &  0  \\
    0  & -1  &  0  &  i  &  0  &  0  \\
    1  &  0  &  i  &  0  &  0  &  0  \end{array}\right).
\end{eqnarray}
Then the transformation matrix $\mathcal{A}$ is as follows:
\begin{equation}
    \mathcal{A} = \frac{1}{\sqrt{6}}
    \left(\begin{array}{cccccc}
    -\sqrt{3}   &  0       &  1 &  0         &  0         & -\sqrt{2} \\
     0          & -1       &  0 &  \sqrt{3}  & -\sqrt{2}  & 0         \\
    -i\sqrt{3}  &  0       & -i &  0         &  0         & i\sqrt{2} \\
     0          & -i       &  0 & -i\sqrt{3} & -i\sqrt{2} &  0        \\
     0          &  2       &  0 &  0         & -\sqrt{2}  &  0        \\
     0          &  0       &  2 &  0         &  0         &  \sqrt{2} \end{array}\right),
\end{equation}
where the $t_{2g}$ orbitals have been treated as a system with $l_{\text{eff}}=1$.

In the natural single particle basis, SOC matrix is transformed into:
\begin{eqnarray}
    H_{soc} = 
    \left(\begin{array}{cccccc}
   -\zeta/2 &  0       &  0       &  0       & 0      & 0        \\
    0       & -\zeta/2 &  0       &  0       & 0      & 0        \\
    0       &  0       & -\zeta/2 &  0       & 0      & 0        \\
    0       &  0       &  0       & -\zeta/2 & 0      & 0        \\
    0       &  0       &  0       &  0       & \zeta  & 0        \\
    0       &  0       &  0       &  0       & 0      & \zeta \end{array}\right).
\end{eqnarray}
Meanwhile the Coulomb interaction term is transformed as:
\begin{equation}
    \sum_{\alpha\beta\delta\gamma}U_{\alpha\beta\delta\gamma} 
    d_{\alpha}^{\dagger}d_{\beta}^{\dagger} d_{\delta}d_{\gamma}
  = \sum_{mnkl} \tilde{U}_{mnkl} c_m^\dagger c_n^\dagger c_k c_l,
\end{equation}
with 
\begin{equation}
   \tilde{U}_{mnkl} = \sum_{\alpha\beta\delta\gamma} U_{\alpha\beta\delta\gamma}
   \mathcal{A}_{m\alpha}^{\dagger} \mathcal{A}_{n\beta}^{\dagger} \mathcal{A}_{\delta k} \mathcal{A}_{\gamma l}.
\end{equation}
\begin{figure}[ht]
\centering
\includegraphics[scale=0.80,angle=-0]{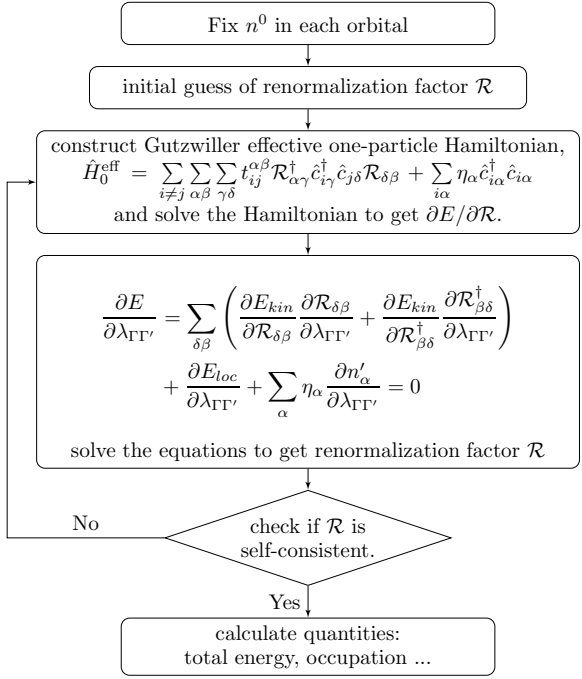}
\caption{Flowchart of the RIGA self-consistent loop to minimize total 
energy $E(n^0)$ with respect to $|\Psi_0\rangle$ and $\lambda_{\Gamma\Gamma'}$.
\label{flow}}
\end{figure}

In this paper, we define expectation value with uncorrelated wave function:
\begin{equation}
    O^{0}=\langle\Psi_{0}|\hat{O}|\Psi_{0}\rangle,
\end{equation}
while expectation value with Gutzwiller wave function is defined as:
\begin{equation}
    O=O^G=\langle\Psi_{G}|\hat{O}|\Psi_{G}\rangle.
\end{equation}
During the minimization process, two following  constraints are forced,
\begin{equation}
\langle\Psi_{0}|\mathcal{P}^{\dagger}\mathcal{P}|\Psi_{0}\rangle=1,
\end{equation}
and 
\begin{equation}
\langle\Psi_{0}|\mathcal{P}^{\dagger}\mathcal{P}n_{i\alpha}|\Psi_{0}\rangle
=\langle\Psi_{0}|n_{i\alpha}|\Psi_{0}\rangle.
\end{equation}

In the present paper, the second constraint is satisfied in the following way. We first calculate
the total energy of the trial wave function with both the left-hand  and right-hand side of the above equation 
equaling to some desired occupation number $n^0_{\alpha}$. Then  we minimize the energy
respect to $n^0_{\alpha}$ at the last step.

The remaining task is to minimize the variational ground energy $E=E_{kin}+E_{loc}$
with respect to $\lambda_{\Gamma\Gamma'}$ and $|\Psi_{0}\rangle$, 
and fulfill the previous two constraints. Here, 
\begin{equation}
E_{kin}=\langle\Psi_G|H_{kin}|\Psi_G\rangle = 
\sum_{ij}\sum_{\gamma\delta}\tilde{t}_{ij}^{\gamma\delta}
\langle\Psi_0| c_{i\gamma}^{\dagger}c_{j\delta}|\Psi_0\rangle,
\end{equation}
and 
\begin{equation}
E_{loc} = \langle\Psi_G|H_{loc}|\Psi_G\rangle = \text{Tr}(\phi^{\dagger}{H}_{loc}\phi),
\end{equation}
with $\tilde{t}$, $\mathcal{R}$ and $\phi$ defined as:
\begin{equation}
   \tilde{t}_{ij}^{\gamma\delta}=\sum_{\alpha\beta}t_{ij}^{\alpha\beta}
   \mathcal{R}_{\alpha\gamma}^{\dagger}\mathcal{R}_{\delta\beta}
\end{equation}
\begin{equation}
\mathcal{R}_{\alpha\gamma}^{\dagger}
=\frac{\text{\text{Tr}\ensuremath{\left(\phi^{\dagger}{c}_{\alpha}^{\dagger}\phi{c}_{\gamma}\right)}}}
      {\sqrt{n_{\gamma}^{0}(1-n_{\gamma}^{0})}}, 
\end{equation}
\begin{equation}
\phi_{II'}=\langle I|\mathcal{P}|I'\rangle\sqrt{\langle\Psi_{0}|I'\rangle\langle I'|\Psi_{0}\rangle},
\end{equation}
where $|I\rangle$ ($|I'\rangle$) stands for a many-body Fock state and $n_{\gamma}^0 = \langle \Psi_0| n_{\gamma}|\Psi_0 \rangle$.

The flowchart of RIGA method is shown in Fig.\ref{flow}.
For fixed $n^{0}_{\alpha}$ in each orbital, 
minimizing $E$ with respect to $|\Psi_0\rangle$ and $\lambda_{\Gamma\Gamma'}$
can be divided into two steps in each iterative process. 
Firstly,
fix Gutzwiller variational parameters $\lambda_{\Gamma\Gamma'}$ and find optimal uncorrelated wave 
function by solving effective single particle Hamiltonian,
\begin{equation}
    {H}_{0}^{\text{eff}}=\sum_{i\ne j}\sum_{\gamma\delta}
   \tilde{t}_{ij}^{\gamma\delta}{c}_{i\gamma}^{\dagger}{c}_{j\delta}
   + \sum_{i\alpha}\eta_{\alpha}{c}_{i\alpha}^{\dagger}{c}_{i\alpha},
\end{equation}
where Lagrange parameters $\eta_{\alpha}$ are used to minimize the
variational energy fulfilling Gutzwiller constraints. 
Secondly, we
fix the uncorrelated wave function, and optimize the variational energy
with respect to Gutzwiller variational parameters $\lambda_{\Gamma\Gamma'}$,
\begin{align}
\frac{\partial E}{\partial\lambda_{\Gamma\Gamma'}} & = \sum_{\delta\beta}\left(
\frac{\partial E_{kin}}{\partial\mathcal{R}_{\delta\beta}}
\frac{\partial\mathcal{R}_{\delta\beta}}{\partial\lambda_{\Gamma\Gamma'}} + 
\frac{\partial E_{kin}}{\partial\mathcal{R}_{\beta\delta}^{\dagger}}
\frac{\partial\mathcal{R}_{\beta\delta}^{\dagger}}{\partial\lambda_{\Gamma\Gamma'}} \right) \nonumber\\
& + \frac{\partial E_{loc}}{\partial\lambda_{\Gamma\Gamma'}} + 
\sum_{\alpha}\eta_{\alpha}\frac{\partial n_{\alpha}'}{\partial\lambda_{\Gamma\Gamma'}}=0,
\end{align}
where $n_{\alpha}'=\langle\Psi_{0}|\mathcal{P}^\dagger \mathcal{P} {n}_{\alpha}|\Psi_{0}\rangle$.
In this way, $\lambda_{\Gamma\Gamma'}$
and $|\Psi_{0}\rangle$ are self-consistently determined.

For the fix $n^0_{\alpha}$ algorithm, we need to scan the $n^0_{\alpha}$  to get the global ground state of the studied system.
In this paper, because SOC will split the $t_{2g}$ orbitals into two fold $j_{\text{eff}}=1/2$ and four 
fold $j_{\text{eff}}=3/2$ states, we can introduce an alternative variable $\delta n^{0}$ to determine $n^0_{\alpha}$ for each orbital.
The occupation polarization $\delta n^{0}$ is defined as:
\begin{equation}
   \delta n^0 = n_{3/2}^0 - n_{1/2}^0, 
\end{equation}
in which $n_{3/2}^0$ and $n_{1/2}^0$ stand for the average occupation number of 
lower ($j_{\text{eff}} = 3/2$) and 
upper ($j_{\text{eff}} = 1/2$) orbitals respectively. Since total electron number of the system is 
fixed to be $4n_{3/2}^0 + 2n_{1/2}^0 = 4$, we have $0 \leq \delta n^0 \leq 1$.
$\delta n^{0}$ ($n^0$) corresponding to ground state is denoted by $\delta n^0_g$ ($n^0_g$).

In the present paper, we also use DMFT+CTQMC method to crosscheck our results derived by RIGA. 
For DMFT+CTQMC method,  
the system temperature is set to be $T=0.025$ (corresponding to inverse temperature $\beta=40$). 
In each DMFT iteration, typically 
$4\times 10^8$ QMC samplings have been performed to reach sufficient numerical accuracy\cite{Huang:PhysRevB.86.035150}.

\section{Results and discussion}
\label{sec:results}

%
\subsection{$U$-$\zeta$ phase diagram}
\label{subsec:rot}

\begin{figure}[ht]
\centering
\includegraphics[scale=0.65,angle=0]{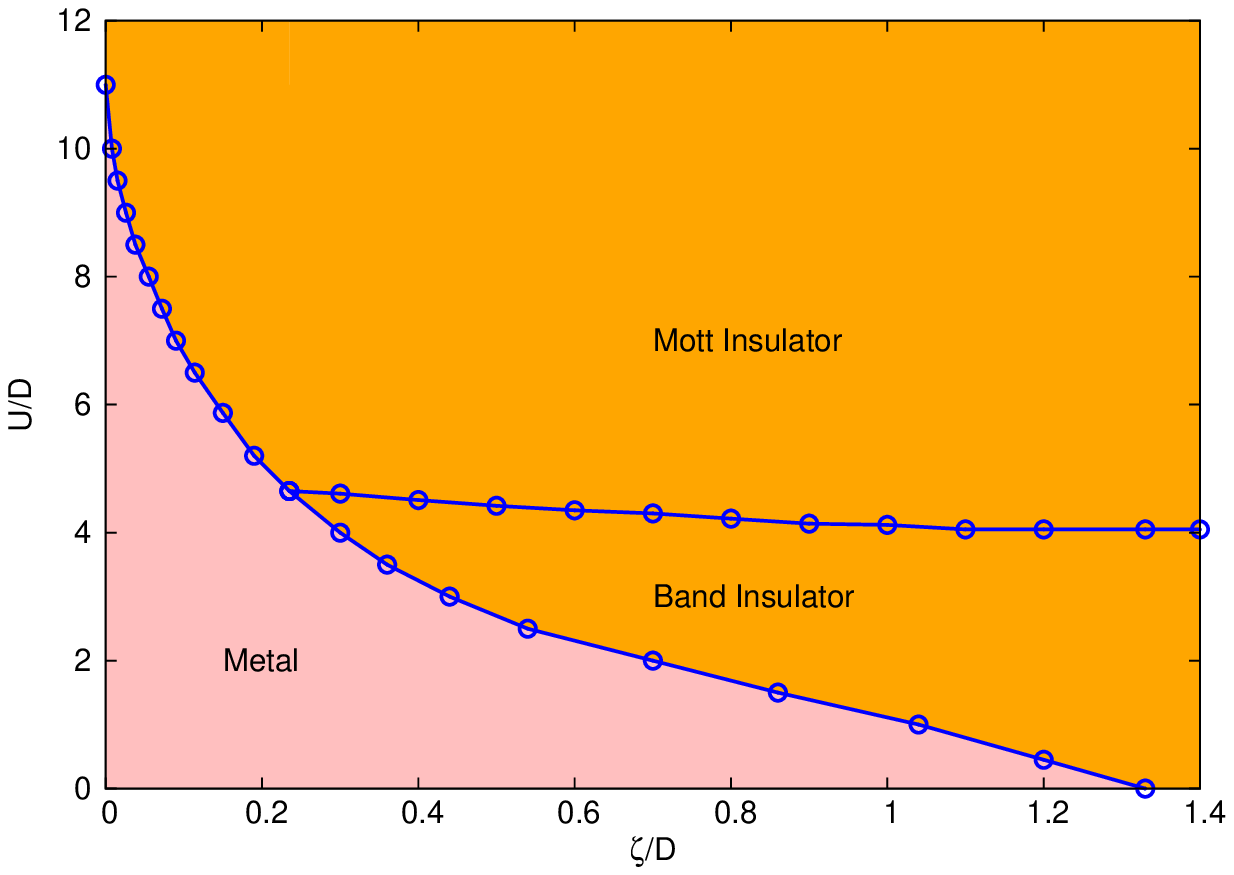}
\includegraphics[scale=0.65,angle=0]{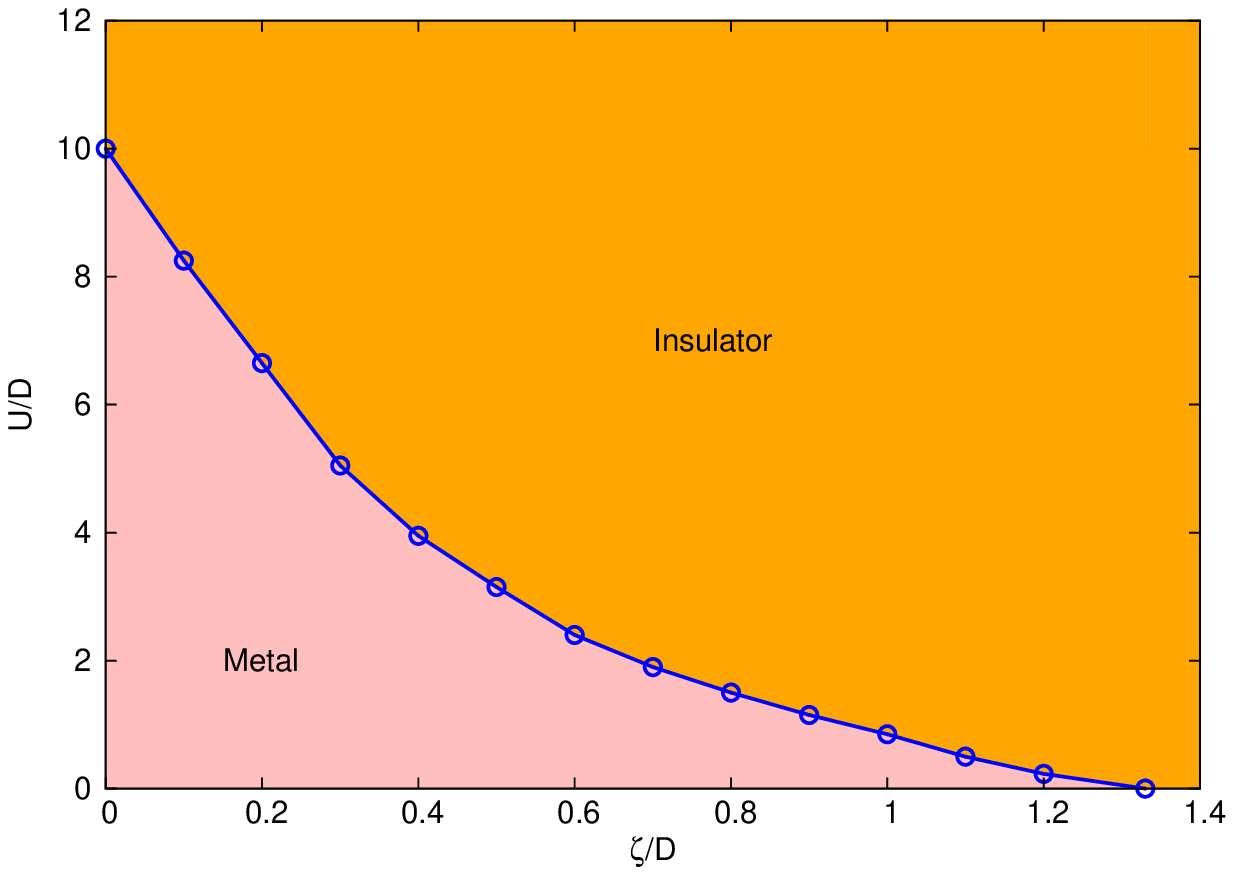}
\caption{(Color online) Phase diagram of three-band Hubbard model with full Hund's coupling terms 
in the plane of Coulomb interaction $U(J/U=0.25)$ and spin-orbit coupling $\zeta$. 
Upper panel: The phase diagram is calculated by RIGA at zero temperature. Lower panel: 
The phase diagram is 
calculated by DMFT+CTQMC with finite temperature $T=0.025$.}
\label{ph1}
\end{figure}

In this subsection, we mainly focus on phase diagram for the three-band 
model proposed in Eq.(\ref{eq1}).
The obtained $U-\zeta$ phase diagrams with $J/U = 0.25$ are shown in Fig.\ref{ph1}.
The upper panel shows the phase diagram calculated by zero
temperature RIGA method, while the calculated results by DMFT+CTQMC method at finite temperature 
is shown in the lower panel.  The results obtained by two different methods are consistent with 
each other quite well except that DMFT+CTQMC can not 
distinguish between band insulator and Mott insulator, which will be explained later.
Apparently, there exists three different phases in $U-\zeta$ plane: 
metallic state in the lower left corner, 
band insulator in the lower right region and Mott insulator in the upper right region.
The general shape of the phase diagram can be easily understood by considering two limiting cases:
(i) For $\zeta = 0$, one has a degenerate three-band Hubbard model 
populated by $4$ electrons per site. 
The model will undergo an 
interaction driven Mott transition at critical $U_c/D \sim 11.0$ with each band filled 
by $4/3$ electrons. 
(ii) For non-interacting case ($U$ = 0.0), the model is exactly soluble. The three bands are 
degenerate and filled by $4/3$ electrons at $\zeta$=0.0. 
Finite $\zeta$ will split the three degenerate bands into a (lower) $j_{\text{eff}}=3/2$ quadruplet and (upper) 
$j_{\text{eff}}=1/2$ doublet with energy separation being $1.5\zeta$. Increasing $\zeta$ 
will pump electrons from upper to lower orbitals until the upper bands are completely empty and the system undergoes
a metal to band insulator transition, which is expected at $\zeta/D = 1.33$.
%
%
\begin{figure}
\centering
\includegraphics[scale=0.65]{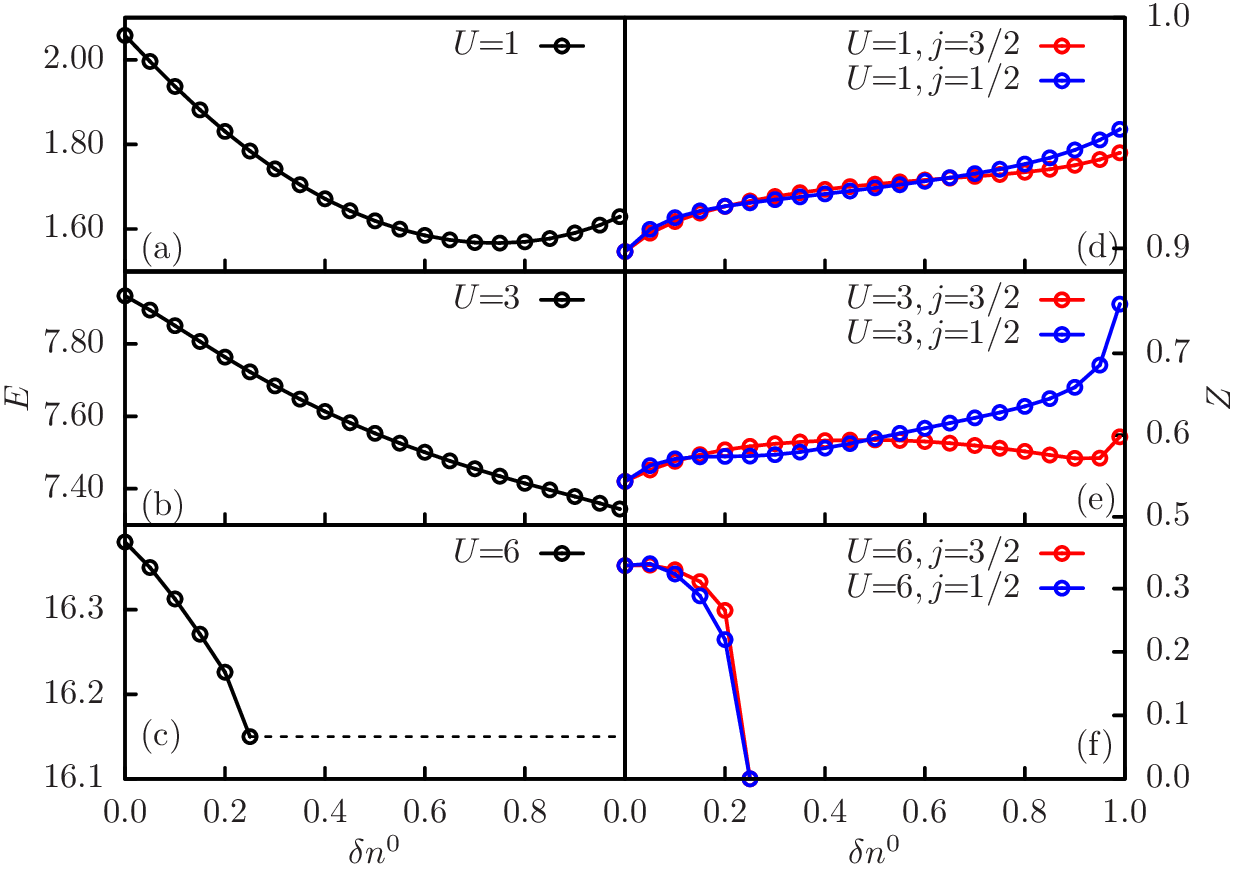}
\caption{(Color online) 
Total energy $E(\delta n^0)$ and quasiparticle weight $Z(\delta n^0)$ as a function of 
occupation polarization $\delta n^0 = n_{3/2}^0 - n_{1/2}^0$ for different values of interaction strength 
$U/D\!=\!1,\ 3,\ 6$ ($J/U\!=\!0.25$) at fixed $\zeta/D = 0.7$ and zero temperature, where $n_{3/2}^0$ is 
average occupation number of $j_{\text{eff}}=3/2$ quadruplet, $n_{1/2}^0$ is average occupation number of 
$j_{\text{eff}} =1/2$ doublet.}
\label{ez-n}
\end{figure}

In order to clarify the way we determine the metal, band insulator, and Mott insulator phases by RIGA method, in Fig.\ref{ez-n}
we plot the total energy and quasiparticle weight as a function of $\delta n^{0}$ defined in the previous section
, where the SOC strength is fixed at $\zeta/D=0.7$, and from top to bottom the Coulomb 
interaction is $U/D\!=\!1.0,\ 3.0,\ 6.0$. 
The ground state of the system is  the state with the lowest energy respect to $\delta n^0$. 
The typical solution for the metal phase is shown in  Fig.\ref{ez-n}a, where the energy minimum occurs 
at $0<\delta n^0_g < 1.0$ corresponding to the case that all orbitals being partially occupied. 
While for a band insulator, as shown in Fig.\ref{ez-n}b as a typical situation, the energy minimum
happens at $\delta n^0_g = 1.0$ corresponding to the case that the $j_{\text{eff}}=3/2$ 
orbitals are fully occupied and $j_{\text{eff}}=1/2$ orbitals are empty, and more over the quasiparticle weight
$Z$ keeps finite when $\delta n^0$ approching unit. And finally the situation of a Mott insulator is illustrated 
in  Fig.\ref{ez-n}c, where the quasiparticle
weight $Z$ vanishes at some critical $\delta n^0$, above which the system is in Mott insulator phase and can no longer
be described by the Gutzwiller variational method.

\begin{figure}[ht]
\centering
\includegraphics[scale=0.65]{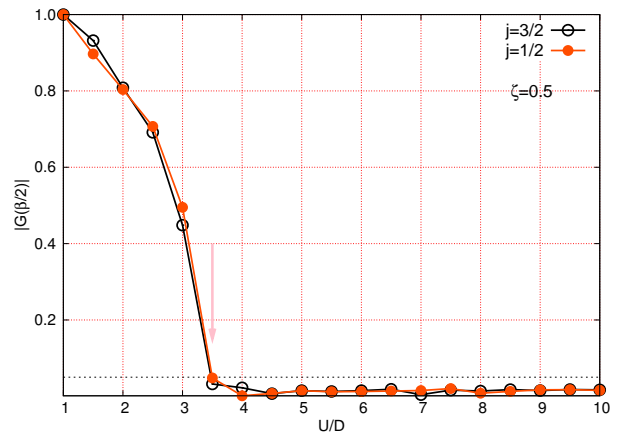}
\caption{(Color online) The imaginary-time Green function at $\tau = \beta/2$ as a function of Coulomb interaction strength $U$.
The SOC strength $\zeta$ is chosen to be 0.5 as a illustration. The calculation is done by DMFT+CTQMC
method at $\beta = 40$. In this figure the normalized quantities by $G(\beta/2)$ at $U/D = 1.0$ are shown
and the arrows correspond to phase transition points.
\label{Gbeta}}
\end{figure}

While in the DMFT+CTQMC calculations, the phase boundary between metal and insulator is identified by measuring
the imaginary-time Green function at $\tau = \beta/2$\cite{Liebsch:PhysRevLett.91.226401,Huang:PhysRevB.85.245110}.
Since $G(\beta/2)$ can be viewed as
a representation of the integrated spectral weight within a few $k_{B}T$ of $E_{F}$, so it can be
used as an important criterion to judge whether metal-insulator transition occurs. The corresponding
results for SOC strength $\zeta/D=0.5$ are shown in Fig.\ref{Gbeta}. Clearly seen in this figure, the critical $U_{c}$ is about 3.5, and
both the $j_{\text{eff}}=3/2$ and $j_{\text{eff}}=1/2$ orbitals undergo metal-insulator transitions simultaneously.
Since there is chemical potential ambiguity in the insulator phase, it is difficult to further distinguish Mott insulator from band 
insulator by DMFT+CTQMC method. Therefore we only calculate the phase boundary between the metal and
insulator phase, which is  in good agreement with the results obtained by RIGA.

Now, we come back to discussed the phase diagram obtained by RIGA and DMFT.
When both the Coulomb interaction $U$ and SOC $\zeta$ are finite, 
the phase diagram looks a bit complicated. By considering different values of SOC, we divide the 
phase diagram vertically into three regions. 

Firstly,  for $0.00< \zeta/D < 0.24$, 
with increasing Coulomb interaction $U$, our calculation by RIGA predicts
a transition from metal to Mott insulator. 
The transition is characterized by the vanishing of quasiparticle weight as discussed previously.
The critical Coulomb interaction $U_c$ decreases drastically with the increment of SOC, 
the effect of SOC tends to enhance the Mott MIT greatly. The DMFT results show very
similar behavior in this region as shown in lower panel of Fig.\ref{ph1}, except that the  $U_c$ 
obtained by DMFT has weaker dependence on the strength of  SOC compared to 
that of RIGA. From the view point of DMFT, the suppression of  the metallic phase by SOC can be
explained quite clearly. For the effective impurity model in DMFT, the metallic phase corresponds 
to a solution when the local moments on the impurity site are fully screened by the electrons in heatbath
through the Kondo like effect. With the SOC, there is an additional channel to screen the local spin moment 
other than the Kondo effect, which leads to the formation of spin-orbital singlets. This additional screening channel, 
which is completely local, will thus compete with the Kondo effect and suppress the metallic solution. There is
no net local moment left in this type of Mott phase, and the ground state is simply a product state of local
spin-orbital singlets on each site.
 
For $0.24 < \zeta/D < 1.33$, two successive phase transitions 
are observed with the increment of $U$. The transition from metallic to band insulator phases occurs firstly, and
followed by another transition to the Mott insulator phase. In the intermediate $U$ region, the effective band
width of system is reduced by the correlation effects, which drives the system into a band insulator phase with
relative small band splitting induced by SOC. Further increasing interaction strength $U$ will push the
system to the Mott limit. Although this process is believed to be a crossover rather than a sharp phase transition,
our RIGA calculation provides a mean field description for these two different insulators, where in the 
band insulator phase the interaction effects only renormalize the effective band structure and do not suppress
the coherent motion of the electrons entirely. Similar behavior can also be obtained by DMFT method, where 
the quasiparticle weight determined by DMFT selfenergy keeps finite for the band insulator phase and vanishes 
for the Mott insulator phase.

At last, for $\zeta/D > 1.33$ region, the orbitals are fully polarized with 
electrons fully occupied $j_{\text{eff}}=3/2$ bands and fully empty $j_{\text eff}=1/2$ bands at $U=0.0$, 
indicating that the system is in the band insulator state already in the non-interacting case. 
Similar band insulator to Mott insulator transition will be induced with further increment of $U$ in the RIGA description, 
as discussed before.

\begin{figure}[ht]
\centering
\includegraphics[scale=0.65]{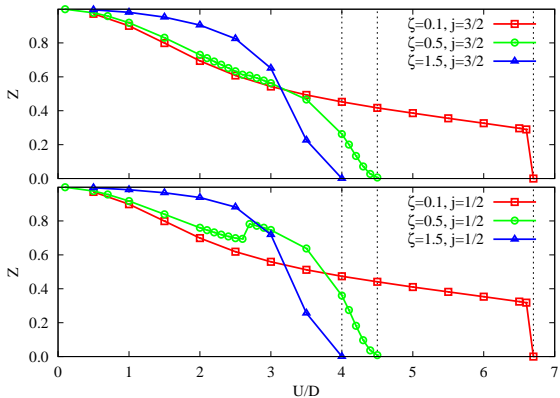}
\caption{(Color online) Quasiparticle renormalization factors $Z$ of the lower orbitals ($j_{\text{eff}}=3/2$ quadruplet) 
and upper orbitals ($j_{\text{eff}}=1/2$ doublet) 
as function of Coulomb interaction $U(J/U=0.25)$ for different values of SOC ($\zeta/D=0.1,0.5,1.5$). The dashed lines 
label the critical $U$ for transition to Mott state.
The results are obtained by zero temperature RIGA method.
\label{J1_ZU}}
\end{figure}
\begin{figure}[ht]
\centering
\includegraphics[scale=0.65]{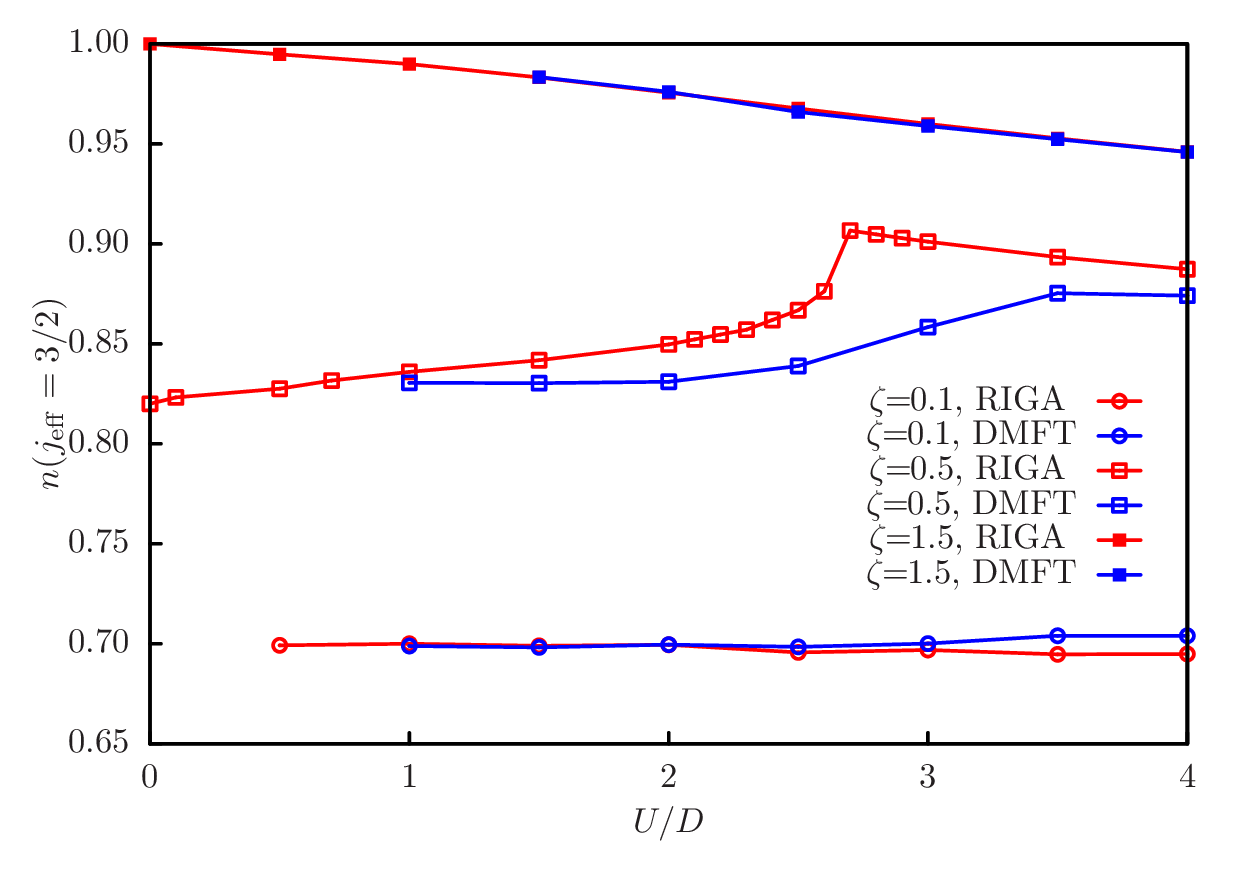}
\caption{(Color online) Occupation number of the lower orbitals ($j_{\text{eff}}=3/2$ quadruplet) with 
increasing Coulomb $U(J/U=0.25)$ for selected SOC ($\zeta/D=0.1,0.5,1.5$). Both the calculated results  
by RIGA and DMFT(CTQMC) are presented.
\label{J1_NU}}
\end{figure}
\begin{figure}[ht]
\centering
\includegraphics[scale=0.65]{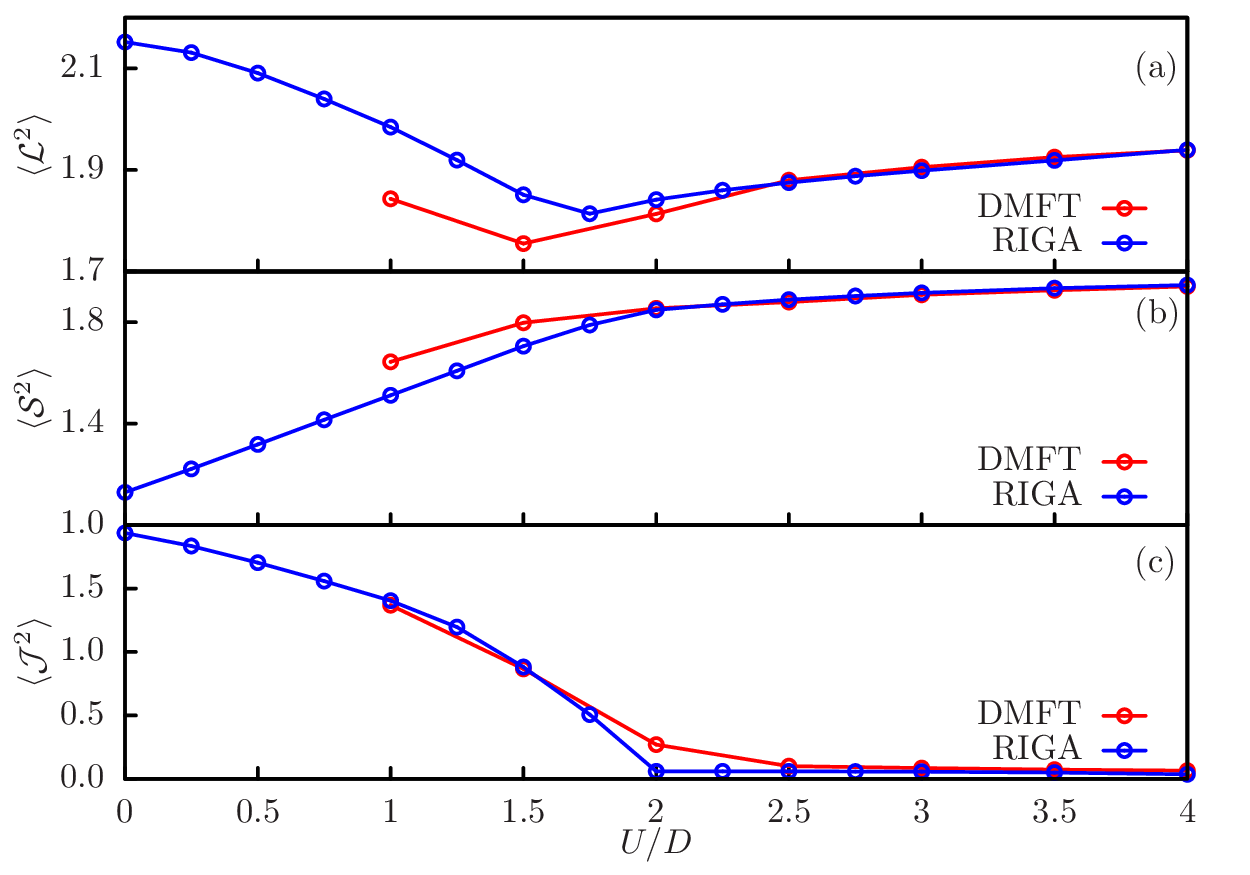}
\caption{(Color online) Expectation value of orbital angular momentum $\langle \mathcal{L}^2 \rangle$, 
spin angular momentum $\langle \mathcal{S}^2 \rangle$, and total angular momentum 
$\langle \mathcal{J}^2 \rangle$ as function of Coulomb interaction $U$ with fixed 
spin-orbit coupling strength $\zeta/D=0.7$. It is derived by RIGA at zero temperature and 
DMFT+CTQMC at $\beta=40$ respectively.
\label{lsj}}
\end{figure}

For several typical SOC parameters ($\zeta/D = 0.1, 0.5, 1.5$) in the three regions defined above,
we study the evolutions of quasiparticle weight and band specific occupancy with Coulomb interaction.
The quasiparticle weight for selected SOC with 
increasing $U$ is plotted in Fig.\ref{J1_ZU}. The upper (lower) panel shows the quasiparticle weight for 
$j_{\text{eff}}=3/2$ ($1/2$) orbitals.
Note the quasiparticle weight in RIGA is defined as the eigenvalues of the Hermite matrix $\mathcal{R}^{\dagger}\mathcal{R}$.
For both $\zeta/D=0.1$ and $1.5$, the quasiparticle weights decrease from 1 to 0 monotonically with the increasing interaction strength
$U$ and $J$ until the transition to Mott insulator phase. 
While for $\zeta/D=0.5$, there exists 
a kink at $U/D=2.7$  in the lower panel ($j_{\text{eff}}=1/2$), which corresponds to the transition from metal to band insulating state. 
For transition to Mott insulating state, quasiparticle weights for all the orbitals reach zero simultaneously, 
with $U_c/D$ = 6.7 for $\zeta/D$ = 0.1,  $U_c/D$ = 4.5 for $\zeta/D$ = 0.5, and $U_c/D$ = 4.0 for $\zeta/D$ = 1.5.

The occupation number of the (lower) $j_{\text{eff}}=3/2$ orbitals as a function of on-site Coulomb 
interaction $U$
is ploted in Fig.\ref{J1_NU} for three typical SOC strength.
For $\zeta/D = 0.1$, to some extent, the occupation behavior 
is similar to $\zeta = 0$ case, in which the occupation number is only slightly
changed by the interaction.
The situation is quite different for $\zeta/D = 0.5$, where the occupation of the $j_{\text{eff}}=3/2$ orbital
increases with interaction at the beginning and decreases slightly after the transition to the band insulator phase.
The non-monotonic behavior here is mainly due to the competition between the repulsive interaction $U$ and Hund's rule
coupling $J$. The effect of $U$ will always enhance the splitting of the local orbitals 
to reduce the repulsive interaction among these orbitals. While the Hund's rule coupling intents to distribute the electrons
more evenly among different orbitals.
For $\zeta/D=1.5$ case, occupation number in the two subsets is fully polarized at $U=0$ and the effect of Hund's coupling term
will reduce the occupation of the $j_{\text{eff}}=3/2$ orbital monotonically.

At last, the expectation value of the total orbital angular momentum $\langle \mathcal{L}^2 \rangle$, 
spin angular momentum  $\langle \mathcal{S}^2 \rangle$ and 
total angular momentum  $\langle \mathcal{J}^2 \rangle$ as a function of Coulomb interaction $U$ are 
plotted in Fig.\ref{lsj}, where $\zeta/D$ is fixed to $0.70$. 
In the non-interacting case, all the three expectation values can be calculated exactly and they
will approach the atomic limit with the increment of interaction $U$ and $J$. In the atomic limit the 
SOC strength  is much weaker than the Hund's coupling $J$, the ground state can be
well described by the $LS$ coupling scheme, where the four electrons will first form a state with 
total orbital angular momentum $\mathcal{L}=1$ and total spin momentum $\mathcal{S}=1$, and then form a spin-orbital 
singlet state with total angular momentum $\mathcal{J}=0$. From Fig.\ref{lsj}, we can find that the system approaches 
the spin-orbital singlet quite rapidly after the transition to the band insulator phase.

%
 \begin{figure}[ht]
 \centering
 \includegraphics[scale=0.65]{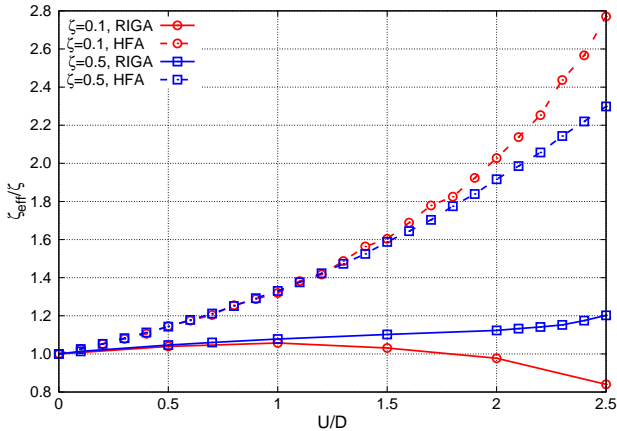}
 \caption{(Color online) Evolution of effective SOC strength with increasing Coulomb $U(J/U=0.25)$ for 
 selected SOC ($\zeta/D=0.1,0.5$), A comparison of results derived by RIGA and HFA are presented.
 \label{J1_LU}}
 \end{figure}

\subsection{Effective spin-orbit coupling}
\label{subsec:soc}
In the multi-orbital system, the interaction  effects will mainly cause two consequences for the metallic phases: 
(1) It will  introduce renormalization factor for the energy bands; 
(2) It will modify the local energy level for each orbital which splits the bands.
For the present model, the second effect will modify the effective SOC, which is another 
very important problem for the spin orbital coupled correlation system. Within the Gutzwiller variational 
scheme used in the present paper, the effective SOC can be defined as:
\begin{equation}
   \zeta_{\text{eff}} = - \frac{1}{2}
   \frac{\partial E_{int}(\delta n^{0})}{\partial \delta n^{0}}  - \frac{1}{2}
   \frac{\partial E_{soc}(\delta n^{0})}{\partial \delta n^{0}},
\end{equation}
where $E_{int}$ and $E_{soc}$ are the ground state expectation values of interaction and SOC terms in the Hamiltonian respectively. 
Note the second term is different from the bare SOC $\zeta_0$ unless $n_\alpha$ is a good quantum number.
If the interaction energy is treated by Hartree Fock mean field approximation (HFA), the above equation gives 
$\zeta_{\text{eff}} = -\partial E^{HF}_{int}(\delta n^{0}) / (2\partial \delta n^{0})+\zeta_0$, 
which will always greatly enhance the spin-orbital splitting with the increasing $U$ as found by some works based
on LDA+U method\cite{PhysRevLett.101.026408,Subedi:PhysRevB.85.020408}. In this section, we compare the results obtained by RIGA and HFA. As shown in 
Fig.\ref{J1_LU}, the effective SOC obtained by HFA increases quite rapidly with the interaction $U$ and $J$. While the results obtained
by RIGA show very different behavior. For weak SOC strength, i.e. $\zeta/D=0.1$, the effective SOC obtained by RIGA increases first then 
decrease. This interesting non-monotonic behavior reflects the competition between the repulsive interaction $U$, 
which intends to increase the occupation difference for $j_{\text{eff}}=1/2$ and $j_{\text{eff}}=3/2$ orbitals,
and  the Hund's rule coupling $J$ , which intents to decrease the occupation difference. While for relatively strong SOC strength, the 
effective SOC increases with interaction $U$ (and $J$) monotonically all the way to the phase boundary indicating the repulsive
interaction $U$ plays a dominate role here. But compared to HFA, the enhancement of effective SOC induced by the interaction is 
much weaker even for the latter case. This is mainly due to the reduction of the high energy local configurations in the Gutzwiller variational
wave function compared to the Hatree Fock wave function, which greatly reduces the interaction energy and its derivative to the orbital occupation.

\section{concluding remarks}

\label{sec:conclusion}
The Mott MIT in three-band Hubbard model with full Hund's rule coupling and SOC is studied 
in detail using RIGA and DMFT+CTQMC methods. 
First, we propose the phase diagram with the strength of electron-electron interaction and SOC.
Three different phases have been found in the $U-\zeta$ plane, which are
metal, band insulator and Mott insulator phases.
For $0.00<\zeta/D<0.24$, increasing Coulomb interaction will induce a MIT transition from metal to Mott insulator.
For $0.24<\zeta/D<1.33$, effect of $U$ will cause two successive transitions, first from metal to band insulator, then to Mott insulator.
For $     \zeta/D>1.33$, a transition from band insulator to Mott insulator is observed.
From the phase diagram,  we find that the critical interaction strength $U_c$ is strongly reduced by the presence of SOC, 
which leads to the conclusion that the SOC will greatly enhance the strong correlation effects in these systems.
Secondly, we have studied the effect of electron-electron interaction on the effective SOC. 
Our conclusion is that the enhancement of effective SOC found in HFA is strongly
suppressed once we go beyond the mean field approximation and include the fluctuation effects by RIGA or DMFT methods.

\section*{acknowledgment}
We acknowledge valuable discussions with professor Y.B. Kim and professor K. Yamaura, and financial support from the 
National Science Foundation of China and that from the 973 program under Contract No.2007CB925000 
and No.2011CBA00108. The DMFT + CTQMC calculations 
have been performed on the SHENTENG7000 at Supercomputing Center of Chinese Academy of 
Sciences (SCCAS).

\bibliography{Hub-soc}

\end{document}